\documentclass[aps,pre,twocolumn,amsmath,amssymb,showpacs,amsfonts]{revtex4}
\usepackage{epsfig}
\usepackage{graphicx}
\usepackage{dcolumn}
\usepackage{bm}
\usepackage{amsmath}

\begin{document}
\title{Finite-time collapse and soliton-like states in the dynamics of dissipative gases}
\author{Itzhak Fouxon}
\affiliation{Raymond and Beverly Sackler School of Physics and Astronomy,
Tel-Aviv University, Tel-Aviv 69978, Israel}
\date{\today}

\begin{abstract}

A study of the gas dynamics of a dilute collection of the inelastically colliding hard spheres
is presented. When diffusive processes are neglected the gas density blows up in a finite time. The blowup is the mathematical expression for 
one of the possible mechanisms for cluster formation in dissipative gases. 
The way diffusive processes smoothen the singularity has been studied. Exact localized soliton-type solutions of the gas dynamics when 
heat diffusion balances non-linear cooling are obtained.
The presented results generalize previous findings for planar flows.

\end{abstract}

\maketitle

A simple case of a many-body dissipative
system is provided by a gas comprising particles whose collisions are inelastic. A common model considers 
a collection of hard spheres undergoing binary collisions, in which
the normal component of the relative velocity of the particles is reduced by a constant coefficient of restitution 
$0\leq r<1$ (for $r=1$ the elastic gas is recovered).
This model of a granular gas is used to describe fluid-like motion of a dilute granular material - a collection of macroscopic
particles with inelastic interactions \cite{BP,Goldhirsch2}. Inelasticity models energy losses to the particles internal degrees 
of freedom. In this Letter we consider the dynamics of granular gases for $r$ sufficiently close to $1$, as explained below. 

At very small inelasticity ($r\lesssim 1$) the free behavior of a granular gas in a finite volume is very similar to the behavior of 
an ordinary gas. The gas evolution brings it to the spatially homogeneous state, differing from a molecular gas in the fact that the 
temperature decays in time due to inelastic energy losses. The decay
law in this homogenous cooling state (below, the HCS) is a power with an inelastic cooling time-scale $t_c$ \cite{Haff}.
However as the inelasticity $1-r$ is increased a "phase transition" (cf. \cite{Brey0}) occurs, still at $1-r\ll 1$. The HCS becomes unstable and the gas starts developing clusters of particles (the clustering instability) \cite{Goldhirsch1,Goldhirsch3,McNamara1,McNamara2}.

Recently, under the condition $1-r\ll 1$ a derivation of the evolution towards the final state of a dilute granular gas was given 
for particular vessel geometry \cite{MFV}.  Two assumptions were made: that the vessel is a channel
and the channel length
is not too large so that the sound travel time through the channel $t_s$ is much smaller than $t_c$. In this geometry only longitudinal excitations of the gas arise which depend on the coordinate that measures distances in the long direction of the channel (the microscopic motions being in $d=2$ or $d=3$ where $d$ is the spatial dimension), while hydrodynamics is one-dimensional. The second assumption implies that sound travels back and forth along the channel many times before inelastic cooling becomes important. As a result, the pressure equilibrates on a time-scale of order $t_s$ while the
processes associated with the inelasticity take place against the background of a spatially homogeneous pressure (cf. \cite{Brey0}),
which varies on the time-scale of the slow process $t_c$. One can describe the slow dynamics using a single scalar equation. A study of this
equation revealed that the non-linear development of the instability is very similar to the gas-liquid
phase transition with many "droplets" arising throughout the gas which eventually coalesce to form a single drop of the new condensed
phase. The final drop is described analytically as an exact soliton-type solution of the full system of gas dynamic equations. This solution
has zero macroscopic velocity and a time-independent density profile. It describes a localized
excitation of the gas. While the analogy between granular gas dynamics and phase ordering was discussed \cite{NoijeErnst1,NoijeErnst2,NoijeErnst3,Baldassarri} and explored numerically
\cite{Puri} before, the theory in \cite{MFV} is
unique in that it provides a rather complete description of the evolution of the gas under the stated assumptions. In particular, it describes analytically the final state of the gas.

The work \cite{MFV} is the most recent in a series of papers \cite{ELM,MP,Fouxon1,Fouxon2,Puglisi,MFV}
devoted to the study of planar motions of a granular gas.
The first study \cite{ELM} introduced one-dimensional gas dynamics as a description of flows arising in a channel and studied
the flows resulting from initial long scale excitations. The studies \cite{MP,Fouxon1,Fouxon2,Puglisi,MFV} were mainly devoted to 
ideal granular gas dynamics (IGGD) obtained by neglecting heat conduction and viscosity. The rationale for studying the IGGD
is that the clustering instability can be described in terms of the competition of two processes - the smoothing of inhomogeneities by
viscous and thermal diffusion processes, having a characteristic time-scale $t_{dif}$, and cooling dynamics corresponding to
{\it ideal}, diffusionless fluid dynamics, whose time-scale is $t_c$. The HCS is unstable when the development of instabilities
by cooling is faster than diffusion processes, $t_c<t_{dif}$. An interesting question is whether the density can grow indefinitely within the IGGD. 
It was found \cite{Fouxon1,Fouxon2,Puglisi} that solutions of the one-dimensional IGGD become singular in a finite time $t_{cr}$.
The density field $\rho$ diverges as $\rho\sim 1/(t_{cr}-t)^2$. Remarkably, the gas pressure $p=\rho T$, where $T$ is the temperature, remains finite at the singularity. The gas freezes at the singularity, $T\sim (t_{cr}-t)^2$, as can be roughly
obtained from $\partial_t T\propto -\rho T^{3/2}$ (see below) by substituting $\rho T=const$. Thus, at least for planar motions,
the IGGD leads to an indefinite growth of the gas density.

Below we show that the finite-time density singularities of the IGGD
carry over to higher dimensions. The analysis is performed in the fast sound regime described above.
Then we show that the soliton-type states exist also in higher dimensions and are not restricted to the fast sound regime.

Consider both $d=2$ (gas of discs in a plane) and $d=3$ (gas of hard spheres in a box) cases. Particles masses are set to unity and their
diameter is designated by $\sigma$. The IGGD equations read (see \cite{Fouxon1,Fouxon2} and references therein)
\begin{eqnarray}&&
\partial_t \rho+\left(\bm v\cdot\bm \nabla\right)\rho=-\rho \nabla\cdot \bm v,
\label{a300} \\&&
\rho\left[\partial_t \bm v+\left(\bm v\cdot\bm \nabla\right)\bm v \right]=
- \nabla p, \label{a311} \\&&
\partial_t T+\left(\bm v\cdot\bm \nabla\right)T=-(\gamma-1) T \nabla\cdot \bm v-\Lambda\rho
T^{3/2},\label{a333}
\end{eqnarray}
where $\bm v$ is the velocity, $\gamma$ is the adiabatic index of the gas ($\gamma=2$ and $5/3$ for $d=2$
and $d=3$, respectively), $\Lambda=2 \pi^{(d-1)/2} (1-r^2) \sigma^{d-1}/[d\,
\Gamma(d/2)]$ (see \textit{e.g.} \cite{Brey}) and $\Gamma(\dots)$ is the
gamma-function. Equations~(\ref{a300})-(\ref{a333}) differ from the
hydrodynamic equations for a dilute gas of elastically colliding
spheres only by the presence of the inelastic cooling term $-\Lambda \rho T^{3/2}$
which is proportional to the product of the average energy loss per collision, $\sim (1-r^2)T$,
and the collision rate, $\sim \rho T^{1/2}$. The system (\ref{a300})-(\ref{a333}) applies when the conditions 
for the local thermal equilibrium are satisfied. Besides the usual conditions for gases \cite{Landau10}, this 
leads to the quasi-elasticity condition $1-r^2\ll 1$. The inequality guarantees that the characteristic cooling time $t_c=2/\Lambda\rho_0^{1/2}p_0^{1/2}$
inferred from Eq.~(\ref{a333})
where $\rho_0$ is the average gas density (the total gas
mass divided by the volume of the channel), and $p_0$ is a characteristic value of the initial pressure,
is much larger than the mean free time $\tau_{mean}\sim 1/(\sigma^{d-1} \rho_0^{1/2}p_0^{1/2})$. 
Then the system (\ref{a300})-(\ref{a333}) describes well the evolution of long-scale excitations with large 
Reynolds number $Re\sim \rho l\sigma^{d-1} v/\sqrt{T}$, cf. \cite{Landau} and Eqs.~(\ref{hydrodynamics2})-(\ref{hydrodynamics3}) 
below. Here $l$ is the excitation scale, $\rho$, $v$, $T$ are characteristic values of the fields, and $Re\gg 1$ allows to neglect the heat conduction and viscosity terms in the equations.
Below we use $p$
instead of $T$ as the second independent thermodynamic variable and employ 
\begin{eqnarray}&&
\partial_t p+\left(\bm v\cdot\bm \nabla\right)p
=-\gamma p \nabla\cdot \bm v-\Lambda\rho^{1/2} p^{3/2}, \label{a322}
\end{eqnarray}
instead of Eq.~(\ref{a333}). 
We will assume small $1-r$ so that
the characteristic time-scale $t_s\equiv L\sqrt{\rho_0/p_0}$, at which the sound traverses the characteristic
vessel size $L$, obeys $t_s\ll t_c$. Still we demand $t_s\gg\tau_{mean}$ so that $L$ is much larger than the mean
free path $l_{mean}\sim 1/\rho_0\sigma^{d-1}$. This leads to the range of vessel sizes $l_{mean}\ll L\ll l_{mean}/(1-r^2)$. In this range the evolution of the gas is as follows: the gas relaxes to the
state of local thermal equilibrium on time-scale $\tau_{mean}$, then it relaxes to the state of a spatially
homogeneous pressure on time-scale $t_s$ and only later, on time-scale $t_c$, the inelasticity becomes
relevant. Thus the cooling dynamics occurs against the background of a spatially constant pressure, $\nabla p=0$.
In other words, the elimination of the fast acoustic modes allows so substitute the Eq.~(\ref{a311})
by $\nabla p=0$
or $p=p(t)$. To see this explicitly, we pass to rescaled variables. We measure time in the units
of $t_c$, coordinate in the units of $L$, and the gas density, pressure and velocity in the units
of $\rho_0$, $p_0$ and $L/t_c$ respectively. Keeping the original notation for the rescaled variables,
we observe that Eq.~(\ref{a300}) does not change, while Eqs.~(\ref{a311}) and (\ref{a322}) become
\begin{eqnarray}
\left(t_s/t_c\right)^2\rho\left[\partial_t \bm v+\left(\bm v\cdot\bm \nabla\right)\bm v \right]=- \nabla p,\\
\partial_t p+\left(\bm v\cdot\bm \nabla\right)p
=-\gamma p \nabla\cdot \bm v-2\rho^{1/2} p^{3/2}. \label{a4}
\end{eqnarray}
The leading order approximation in $t_s/t_c\ll 1$ is obtained by dropping the velocity term in the Euler
equation which results in $\nabla p=0$. Thus the pressure depends on time only, $p=p(t)$. Averaging Eq.~(\ref{a4}) over the volume and using $\bm v\cdot\nabla p=0$ we find
\begin{eqnarray}&&
{\dot p}(t)=-2\langle \rho^{1/2}(t)\rangle p^{3/2}(t), \label{a7}
\end{eqnarray}
where angular brackets stand for the spatial average, $\langle \rho^{1/2}(t)\rangle=\int \rho^{1/2}(\bm x, t)d\bm x/V$
with $V$ the system volume. We used that $\langle \nabla\cdot\bm v\rangle=0$ both for periodic or zero boundary conditions on velocity. Expressing now $\nabla\cdot\bm v$ from Eq.~(\ref{a4}) in terms of $p$ and $\rho$,
we rewrite Eq.~(\ref{a300}) as
\begin{eqnarray}&&
\gamma(p\rho)^{-1/2}\left[\partial_t+
\bm v\cdot\bm \nabla\right]\rho^{1/2}=
\rho^{1/2}-\langle \rho^{1/2}\rangle. \label{a5}
\end{eqnarray}
The above reduction of the original system of Eqs.~(\ref{a300})-(\ref{a322}) has great simplicity: the density
field is driven by the spatial fluctuations of $\rho^{1/2}(\bm x, t)$. It is possible to write down a closed
equation for the average $\langle \rho^{1/2}\rangle$. We note that Eq.~(\ref{a300}) can be written as
$\partial_t\rho^{1/2}+(\bm v\cdot\nabla)\rho^{1/2}=-\rho^{1/2}\nabla\cdot\bm v/2$. Averaging the last equation
over space and using that integration by parts implies $-\langle \rho^{1/2}\nabla\cdot\bm v \rangle =
\langle (\bm v\cdot\nabla)\rho^{1/2}\rangle$, we find $\langle (\bm v\cdot\nabla)\rho^{1/2}\rangle=-2
\partial_t \langle \rho^{1/2}\rangle$. Finally, performing spatial averaging of Eq.~(\ref{a5})
multiplied by $\rho^{1/2}$ and using $\langle \rho\rangle=1$ we obtain
\begin{eqnarray}&&
d\chi/d\tau=-1+\chi^2, \ \ \chi(\tau)\equiv \langle\rho^{1/2}\rangle(\tau), \label{a6}
\end{eqnarray}
where following \cite{MFV} we introduced a new time variable $\tau\equiv \int_0^t p^{1/2}(t')dt'/\gamma$.
The last two equations, though they cannot be derived by the same means as in the one-dimensional case,
have exactly the same form as in that case \cite{MFV}. The solution of Eq.~(\ref{a6}) with the initial
condition $\chi_0=\chi(t=0)$ is
\begin{eqnarray}&&
\chi(\tau)=\frac{\chi_0-\tanh \tau}{1-\chi_0\tanh \tau}.  \label{a9}
\end{eqnarray}
The Cauchy-Schwarz inequality $\langle fg\rangle^2\leq \langle f^2\rangle \langle g^2\rangle$ applied to $f=\rho^{1/2}$ and
$g=1$ gives $\langle \rho^{1/2}\rangle\leq \langle \rho\rangle^{1/2}$. Combined with $\langle \rho\rangle=1$ this implies that $\chi_0\leq 1$ with
equality holding only in the case of spatially homogeneous density. As a result the above expression
for $\chi(\tau)$ becomes negative at some finite time $\tau_0$. Since by definition $\chi(\tau)>0$
this signifies that the solution must break down at $\tau<\tau_0$. Indeed the density evolving
according to Eq.~(\ref{a5}) blows up at $\tau<\tau_0$. To see the blowup we note that
the density ${\tilde \rho}(\tau, \bm a)$ in the (Lagrangian) frame moving with the fluid ${\tilde \rho}(\tau, \bm a)\equiv \rho[\tau, \bm x(\tau, \bm a)]$ with $\partial_{\tau}\bm x(\tau, \bm a)=\bm v(\tau, \bm x(\tau, \bm a)$ obeys (we omit tildes with no ambiguity)
\begin{eqnarray}&&
\rho^{-1/2}(\tau, \bm a)\partial_\tau\rho^{1/2}(\tau, \bm a)=
\rho^{1/2}(\tau, \bm a)-\chi(\tau),
\end{eqnarray}
where $\bm a$ is a label of the trajectory (say the initial coordinate $\bm x(0, \bm a)=\bm a$).
Following \cite{MFV}, we pass to $w\equiv 1/\sqrt{\rho}$, that obeys a linear equation
\begin{eqnarray}&&
\partial_{\tau} w(\tau, \bm a)-\chi(\tau) w(\tau, \bm a)=-1. \label{eq101}
\end{eqnarray}
The above equations are the same as in one-dimensional situation and one can take the solution
from \cite{MFV}. The density in the Lagrangian frame is given by
\begin{eqnarray}&&
w(\tau, \bm a)=\frac{w(0, \bm a)+\chi_0\left[\cosh \tau -1\right]-\sinh \tau}{\cosh \tau-\chi_0\sinh \tau}.
\label{a10}
\end{eqnarray}
The solution describes the formation of density singularity exactly as in $d=1$ situation, see details in \cite{MFV}.
The only difference is that in $d>1$ the knowledge of density in the Lagrangian frame does not allow to recover its
profile in real space, while in $d=1$ this is possible \cite{MFV}. Thus we know that the singularity occurs and we know
its law, but we don't know where in space it occurs. The resolution of this question demands the use of vorticity equation
for the calculation of $\bm v(\bm x, t)$, see \cite{GLM}, and it is beyond the scope of the present work.

Thus we have shown that the evolution of any inhomogeneous initial condition according to Eqs.~(\ref{a300})-(\ref{a333}) produces finite time singularity of the density for vessel sizes
$l_{mean}\ll L\ll l_{mean}/(1-r^2)$. We
expect the finite time blowup holds in the whole range of applicability of hydrodynamics
$l_{mean}\ll L$. The reason is that the system of Eqs.~(\ref{a300})-(\ref{a333}) is hyperbolic so that creation of singularities is determined by spatially localized dynamics that fits the previous analysis.
This viewpoint is also confirmed by the existence of a group of exact solutions for planar flows
that have the same finite-time singularity \cite{Fouxon1,Fouxon2}.
Still further work is needed to establish the fact completely.

Does the heat conduction and viscosity arrest the above singularity? The answer is not obvious due to the vanishing of heat
conductivity and viscosity at a singularity with $T=0$.
Here we show indirectly that the conduction arrests the singularity by displaying exact, valid at any $L$, solutions of equations with heat conduction for which the density is not growing. Diffusive processes add to Eqs.~(\ref{a311}) and (\ref{a322}) the terms describing the viscous momentum transfer and the conduction of heat:
\begin{eqnarray}&&
\rho\left[\partial_t v_i+\left(\bm v\cdot\bm \nabla\right)v_i \right]=
- \nabla_i p+\nu_0\partial_j \left(\sqrt{T}\sigma_{ij}\right)
, \label{hydrodynamics2} \\&&
\partial_t p+\left(\bm v\cdot\bm \nabla\right)p=-\gamma p \nabla\cdot \bm v-\Lambda\rho^{1/2}
p^{3/2} \nonumber\\&&
+(2\kappa_0/3)\nabla^2 T^{3/2}
+(\gamma-1)\nu_0\sqrt{T}\sigma_{ij}\partial_jv_i,\label{hydrodynamics3}
\end{eqnarray}
while Eq.~(\ref{a300}) is unchanged. Above
$\sigma_{ij}=\partial_j v_i+\partial_i v_j-2\delta_{ij}\nabla\cdot\bm v/d$ and it is enough
for our purposes here that the coefficients $\nu_0$ and $\kappa_0$ are of order $1/\sigma^{d-1}$
\cite{BP}. Let us search for the solutions of the above system for which the gas performs no
macroscopic motion, $\bm v\equiv 0$. Then Eqs.~(\ref{a300}) and (\ref{hydrodynamics2}) produce
$\rho(\bm x, t)=\rho(\bm x)$ and $p(\bm x, t)=p(t)$. Further, spatial averaging of Eq.~(\ref{hydrodynamics3})
gives ${\dot p}=-\Lambda p^{3/2}\langle \rho^{1/2}\rangle$ which substitution back in
Eq.~(\ref{hydrodynamics3}) gives the equation defining the stationary profile $\rho(\bm x)$:
\begin{eqnarray}&&
\rho^{1/2}-\langle \rho^{1/2}\rangle-(1/3)\nabla^2 \rho^{-3/2}=0. \label{q2}
\end{eqnarray}
where we measure $\rho$ in the units of $\rho_0$ and distances in the units of $l_{cr}\equiv \sqrt{2\kappa_0/\Lambda\rho_0^2}$. The scale $l_{cr}$ is the scale at which the cooling and the heat
diffusion balance each other and it equals the critical length beyond which spatial perturbations of
the HCS are unstable \cite{Goldhirsch1,McNamara1}. The demand of zero heat flux through the boundary and
$T(\bm x, t)=p(t)/\rho(\bm x, t)$ lead to the Neumann
boundary condition (vanishing normal component of $\nabla \rho$) for Eq.~(\ref{q2}). The equation has inhomogeneous solutions which physics is
the same as in other soliton solutions of non-linear physics (see e. g. \cite{Petv}): diffusion
and non-linearity balance each other. The solutions for $d=1$ were considered in \cite{MFV}. It was
shown there that inhomogeneous solutions of Eq.~(\ref{q2}) exist only for $L>l_{cr}$ (in contrast to the solution $\rho=const$ that exists always). As a result at $L>l_{cr}$
where the HCS is unstable the soliton-type solution becomes the global attractor
of the dynamics \cite{MFV}.  The solutions however exist also for $d>1$. As an example consider the $d=3$ spherically symmetric solution of Eq.~(\ref{q2}) with $\rho=\rho(r)$ obeying $\rho'(0)=\rho'(L)=0$.
The numerical solution of the problem is shown in the Figure. 
\begin{figure}[ht]
\vspace{0.5cm}
\begin{tabular}{cc}
 \epsfxsize=6.0cm  \epsffile{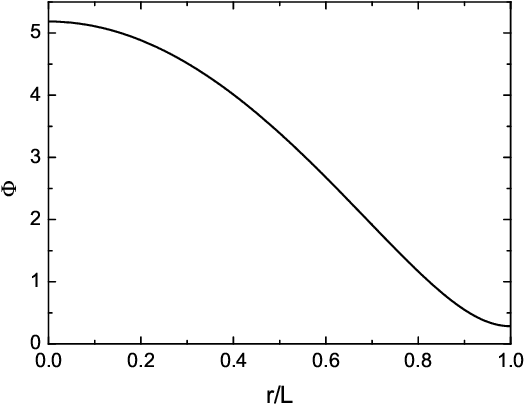}\\
\end{tabular}
\vspace{0.5cm}
\caption{Numerical plot of $\Phi\equiv\langle \rho^{1/2}\rangle^3\rho^{-3/2}$ versus $x=r/L$. The system length $L$ obeys
$L=6l_{cr}$.} \label{eps0.8}
\end{figure}
As in $d=1$ case inhomogeneous solutions
exist only for $L>{\tilde l}_{cr}$ (it can be shown that ${\tilde l}_{cr}$ corresponds to the same length for which the homogeneous cooling state becomes unstable versus spherically symmetric perturbations). Note that $p=\rho T$ is proportional to the energy density since $\rho v^2\equiv 0$. One has
$p=p_0[1+\Lambda\langle \rho^{1/2}\rangle p_0^{1/2}t/2]^{-2}$ both for homogeneous and inhomogeneous density solutions however for inhomogeneous solutions the decay
is slower by $\langle \rho^{1/2}\rangle \leq \langle \rho\rangle^{1/2}$, cf. \cite{MFV}. Thus the transition from the HCS to the soliton state
is the transition to a state with a slower energy decay rate. The above solutions are general and do 
not involve the fast sound assumption.


It is natural to ask if the soliton solutions are global attractors as in $d=1$ case. One can start with the case where sound is the fastest macroscopic process
in the system. Then again the evolution occurs against the background of a spatially uniform pressure.
One finds that thermal conduction effects lead to an additional, Laplacian, term in the effective equation
(\ref{a5}):
\begin{eqnarray}&&
\!\!\!\!\!\!\!\gamma(p \rho)^{-1/2}\!\left[\partial_t\!+\!
\bm v\cdot\bm \nabla\right]\rho^{1/2}\!=\!
\rho^{1/2}\!-\!\langle \rho^{1/2}\rangle\!-\!(1/3)\nabla^2\rho^{-3/2}
. \nonumber
\end{eqnarray}
Clearly soliton solutions are also stationary solutions of the above effective dynamics. In $d=1$
introducing the Lagrangian mass coordinate $m(x, t)=\int_0^x \rho(x', t) dx'$ one may get a closed
description for the evolution of $w(m, t)$. Eq.~(\ref{eq101}) gets modified to (note that $m$ is a
particular case of a Lagrangian label $\bm a$)
\begin{eqnarray}&&
\!\!\!\!\!\partial_{\tau} w(\tau, m)-\chi(\tau) w(\tau, m)=-1+w^{-1}(\tau, m)\partial_m^2w(\tau, m). \nonumber
\end{eqnarray}
It is the study of the above equation that has allowed the complete understanding of the gas evolution in $d=1$ \cite{MFV}.
In $d>1$ such reduction is impossible (derivatives of Lagrangian coordinates
with respect to the initial position are no longer expressible via density only). The study of the above dynamics is 
a subject for future research. 

The above results generalize previous findings for planar hydrodynamic flows \cite{Fouxon1,Fouxon2,MFV} showing 
that the analysis in $d=1$ can serve as a good guide in the study of the fully three (or two) - dimensional hydrodynamics of
granular gases. The non-linear ideal dynamics of free granular gases, obtained by neglecting heat conduction and viscosity, leads to
density divergence in a finite time just like in $d=1$. This singularity provides a robust distinction between the dissipative
and conservative gases (where the density always remains bounded) and describes a possible mechanism for the cluster formation. 
The demonstration was performed in a particular range of vessel sizes and it was argued that the hyperbolic nature of the equations implies that the blowup holds generally. The property that the arising singularities are effectively one-dimensional is known for a simpler gas dynamics where the velocity obeys the Hopf equation $\partial_t\bm v+(\bm v\cdot\nabla)\bm v=0$ and the density obeys the continuity equation.
In both cases, the sign of effective one-dimensionality of singularities is the arising filament structure of the high density regions - the singularities occur in the direction perpendicular to filaments. The filament structure of high density regions in granular gases is observed in \cite{Goldhirsch1}.

We derived new soliton-like states of the gas as exact solutions of the full system of gas dynamic equations. The solutions show
how heat conduction smooths the density singularity described above. They generalize $d=1$ solutions that were shown to describe 
the final state of the gas at least in the fast sound regime \cite{MFV}. Are soliton solutions stable and do they
describe the final state of granular gas in $d=2$ and $d=3$?
Does gas evolution parallel the gas-liquid transition as in $d=1$? These are subjects for further work.

I am very grateful to B. Meerson for numerous fruitful discussions. I am indebted to A. Vilenkin
for help in the numerical solution. I thank I. Goldhirsch for an important discussion. 
R. Sari is acknowledged for
illuminating talks. This work was supported by DIP $0603215013$ and BSF $0603215611$ grants.

\end{document}